%% file: creepycocreator.tex
\documentclass[sigconf]{acmart}

\input{latex/colors}
\input{latex/packages}

\input{latex/statistics}

\input{latex/lists}

\input{meta/terms}
\input{meta/copyright}

\begin{document}
	
\input{meta/title}

\input{meta/teaser}
\input{meta/abstract}

\input{meta/keywords}

\maketitle

\input{sections/introduction}

\input{sections/related_work}
\input{sections/method}
\input{sections/results}

\input{sections/discussion}

\input{sections/limitations} 
\input{sections/conclusion}
	
\input{meta/acknowledgments}
\input{bib/bibliography}

\end{document}

%% file: latex/colors.tex
\usepackage{xcolor}
\usepackage{color}

\definecolor{bananayellow}{rgb}{1.0, 0.88, 0.21}
\definecolor{carrotorange}{rgb}{0.93, 0.57, 0.13}
\definecolor{bleudefrance}{rgb}{0.19, 0.55, 0.91}
\definecolor{armygreen}{rgb}{0.29, 0.33, 0.13}
\definecolor{cadmiumgreen}{rgb}{0.0, 0.42, 0.24}
\definecolor{antiquewhite}{rgb}{0.98, 0.92, 0.84}
\definecolor{lightgray}{rgb}{0.83, 0.83, 0.83}
\definecolor{indianred}{rgb}{0.8, 0.36, 0.36}
\definecolor{ivory}{rgb}{1.0, 1.0, 0.94}
\definecolor{citrine}{rgb}{0.89, 0.82, 0.04}
\definecolor{applegreen}{rgb}{0.55, 0.71, 0.0}

%% file: latex/packages.tex
\usepackage[utf8]{inputenc}
\usepackage{graphicx}
\usepackage{subcaption}

%% file: latex/statistics.tex
\newcommand{\subEtaG}[2]{%
	\ifthenelse{\equal{#1}{\string >.05}}
	{}
	{, $\eta_{G}^{2}=#2$}%
}

\newcommand{\subEta}[2]{%
 \ifthenelse{\equal{#1}{\string >.05}}
 {}
 {, $\eta^{2}=#2$}%
}

\newcommand{\chisq}[3]{$\chi^2(#1) = #2$, $p #3$}
\newcommand{\val}[2]{$\mu = #1$, $\sigma = #2$}
\newcommand{\valS}[2]{$\mu = #1$s, $\sigma = #2$s}
\newcommand{\anova}[4]{$F_{#1, #2}=#3$, $p#4$}

%% file: latex/lists.tex
\usepackage{framed}

\renewenvironment{leftbar}[2][\hsize]
{
    \def\FrameCommand
    {
        {\color{#2}\vrule width 3pt}
        \hspace{0pt}
    }
    \MakeFramed{\hsize#1\advance\hsize-\width\FrameRestore}
}
{\endMakeFramed}

%% file: meta/terms.tex
\newcommand{\ivHigh}{\textsc{highlighting}}
\newcommand{\ivInc}{\textsc{incremental visualization}}
\newcommand{\ivEmb}{\textsc{embodiment}}

%% file: meta/copyright.tex
\copyrightyear{2025}
\acmYear{2025}
\setcopyright{cc}
\setcctype{by-nc}
\acmConference[CHI '25]{CHI Conference on Human Factors in Computing
Systems}{April 26-May 1, 2025}{Yokohama, Japan}
\acmBooktitle{CHI Conference on Human Factors in Computing Systems (CHI
'25), April 26-May 1, 2025, Yokohama,
Japan}\acmDOI{10.1145/3706598.3713720}
\acmISBN{979-8-4007-1394-1/25/04}

%% file: meta/title.tex
\newcommand{\name}{CreepyCoCreator} %
\newcommand{\abbreviation}{\name{}?} %
\title{\abbreviation{} Investigating AI Representation Modes for 3D Object Co-Creation in Virtual Reality}

\author{Julian Rasch}
\orcid{0000-0002-9981-6952}
\affiliation{
  \institution{LMU Munich}
  \city{Munich}
  \country{Germany}  
}
\email{julian.rasch@ifi.lmu.de}

\author{Julia Töws}
\orcid{0009-0001-7369-1426}
\affiliation{%
  \institution{Saarland University,\ Saarland Informatics Campus}
  \city{Saarbrücken}
  \country{Germany}
  }
\email{julia_toews@web.de}

\author{Teresa Hirzle}
\orcid{0000-0002-7909-7639}
\affiliation{%
  \institution{Department of Computer Science, University of Copenhagen}
  \city{Copenhagen}
  \country{Denmark}}
\email{tehi@di.ku.dk}

\author{Florian Müller}
\orcid{0000-0002-9621-6214}
\affiliation{
	\institution{TU Darmstadt}
	\city{Darmstadt}
	\country{Germany}
}
\email{florian.mueller@tu-darmstadt.de}

\author{Martin Schmitz}
\orcid{0000-0002-7332-3287}
\affiliation{%
  \institution{Saarland University,\ Saarland Informatics Campus}
  \city{Saarbrücken}
  \country{Germany}
  }
\affiliation{%
  \institution{Department of Computer Science, University of Copenhagen}
  \city{Copenhagen}
  \country{Denmark}}
\email{mschmitz@cs.uni-saarland.de}

%% file: meta/teaser.tex
\begin{teaserfigure}
	\centering
	\includegraphics[width=\linewidth]{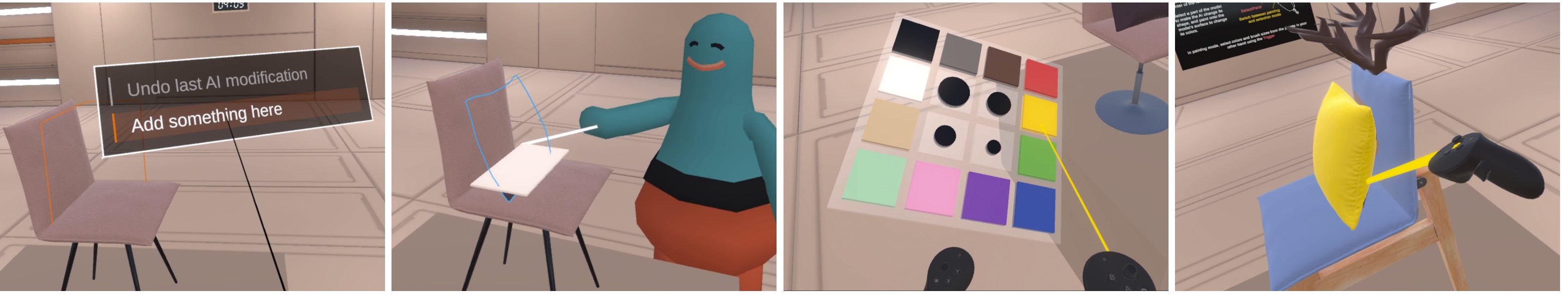}
	\caption{In this paper, we explore VR interaction with an object-generating co-creative AI via a Wizard-of-Oz study on how the AI communicates its intent to users. The study investigates the impact of different modes of AI representation (modification highlighting, incremental visualization of changes, and the embodiment of the avatar) on user perception.}
 \Description{A chair in a VR scene with a hovering interface menu. An embodied AI that adds an object to the chair. A painting palette in VR with different colors and brush sizes. A modified chair with a yellow cushion and a pair of antlers.}
	\label{fig:teaser}
\end{teaserfigure}

%% file: meta/abstract.tex
\begin{abstract}
Generative AI in Virtual Reality offers the potential for collaborative object-building, yet challenges remain in aligning AI contributions with user expectations. In particular, users often struggle to understand and collaborate with AI when its actions are not transparently represented. This paper thus explores the co-creative object-building process through a Wizard-of-Oz study, focusing on how AI can effectively convey its intent to users during object customization in Virtual Reality. Inspired by human-to-human collaboration, we focus on three representation modes: the presence of an embodied avatar, whether the AI’s contributions are visualized immediately or incrementally, and whether the areas modified are highlighted in advance. The findings provide insights into how these factors affect user perception and interaction with object-generating AI tools in Virtual Reality as well as satisfaction and ownership of the created objects. The results offer design implications for co-creative world-building systems, aiming to foster more effective and satisfying collaborations between humans and AI in Virtual Reality.
\end{abstract}

%% file: meta/keywords.tex
\keywords{Co-creative systems; Generative AI; 3D Creation; Virtual Reality; User Studies}

\begin{CCSXML}
	<ccs2012>
	<concept>
	<concept_id>10003120.10003121</concept_id>
	<concept_desc>Human-centered computing~Human computer interaction (HCI)</concept_desc>
	<concept_significance>500</concept_significance>
	</concept>
	<concept>
	<concept_id>10003120.10003121.10003125.10011752</concept_id>
	<concept_desc>Human-centered computing~Haptic devices</concept_desc>
	<concept_significance>500</concept_significance>
	</concept>
	<concept>
	<concept_id>10003120.10003121.10003122.10003334</concept_id>
	<concept_desc>Human-centered computing~User studies</concept_desc>
	<concept_significance>500</concept_significance>
	</concept>
	</ccs2012>
\end{CCSXML}

\ccsdesc[500]{Human-centered computing~Human computer interaction (HCI)}
\ccsdesc[500]{Human-centered computing~User studies}

%% file: sections/introduction.tex
\section{Introduction}
Building 3D worlds and objects for Virtual Reality (VR) has become increasingly accessible, enabling users to immerse themselves in detailed environments and actively (co-)create them. Recent advances have also seen the integration of generative Artificial Intelligence (AI) in this process~\cite{10.1145/3613904.3642579}, leveraging text prompts and other inputs to generate elaborate 3D objects and even entire 3D worlds.
However, these emerging approaches often rely on a "set-it-and-forget-it" approach, where users input a prompt and receive a fully realized 3D object or world with limited opportunities for customization or interaction. 
While this trend toward AI-driven generation can democratize access to high-quality 3D objects, it challenges human agency and shifts control from humans to AI.

To empower users to get more involved in the design process, a stream of research has started exploring co-creative systems, which enable users to collaborate with AI in a creative process \cite{davis_human-computer_2013}. These systems have been applied across a variety of domains, including music, storytelling, game design, and visual arts~\cite{liapis_sentient_2013,lucas_stay_2017,davis_empirically_2016,oh_i_2018,lin_it_2020}. 
There has been growing interest in extending such co-creative systems to 3D object co-creation in recent years. AI systems can now generate and modify 3D objects based on user inputs such as text prompts or sketches~\cite{poole_dreamfusion_2022, lin_magic3d_2023, michel_text2mesh_2021, jun_shap-e_2023, mikaeili_sked_2023, lun_3d_2017}, and some systems even allow for interactive refinement of AI outputs \cite{liu_interactive_2018, urban_davis_designing_2021}.

Despite these advances, a critical question remains largely unexplored: how should the actions of a co-creative tool for object generation be represented in VR? 
This question touches on fundamental aspects of co-creative interactions and raises several essential considerations:
(1) Human collaborative work often involves asynchronous or gradual changes, making immediate modifications potentially counterintuitive. Should AI-driven changes to 3D objects be presented incrementally, mirroring the rhythm of human collaboration, or is immediate modification more appropriate in co-creation?
(2) Drawing from the literature on non-player characters (NPCs), the embodiment is often deemed crucial for effective interaction. However, does this hold for a co-creative system focused on object generation, which may not inherently require an embodied visualization as NPCs do? Does the user perception of the system change from a tool to a collaborator only through embodiment, and if yes, what are the potential downsides?
(3) Decades of HCI research and design practices emphasize highlighting focus or activity (e.g., showing all cursors in collaborative text editing tools or speaker indicators in video conferencing). 
Is visually indicating where and how AI modifications occur sufficient for co-creative object generation systems, or do these systems necessitate new ways of co-creative interaction?

Addressing these questions is pivotal to advancing the design of co-creative systems for 3D objects and world generation.
This paper contributes to answering these questions by investigating different representations of AI actions in the context of co-creating 3D objects in VR (see \autoref{fig:teaser}).
Specifically, we examine the aforementioned three aspects of co-creating a 3D object in VR in a Wizard-of-Oz user study on (1) perception of embodiment, (2) incremental visualization of co-creative contributions, and (3) effect of highlighting the area where the AI is going to perform a modification. 
Our findings show that co-creating with an embodied AI significantly influences the perceived AI contribution to the created model. Further, highlighting does not increase predictability or communication with the AI system but decreases users' enjoyment and a perceived partnership with the AI. Finally, users pay more attention to the AI when it uses an incremental generation.

In summary, the contributions of this paper are two-fold:
First, we conduct a Wizard-of-Oz user study to examine the influence of highlighting, incremental visualization, and embodiment on user perception, collaboration, overall system appeal, and behavioral engagement. We contribute empirical insights into how the different representation strategies affect the co-creative experience in VR-based object-building.
Secondly, based on the study's results, we provide implications for designing future co-creative object-building tools in VR.

%% file: sections/related_work.tex
\section{Related Work}
In the following, we discuss related work in 3D modeling in VR, AI for 3D object generation, co-creative systems in general, embodied AI, and AI contribution visualization.

\subsection{3D Modeling in VR}
Three-dimensional modeling has been a topic of interest in computer research since as early as the 1970s \cite{clark_designing_1976}, with HoloSketch \cite{deering_holosketch_1995} being among the earliest contributions to 3D modeling software for immersive environments. 
In general, there are several ways to approach 3D design in VR. For instance, sketching, though intuitive and quick, suffers from a lack of haptic feedback mid-air. Some systems mitigate inaccuracies by smoothing out hand-drawn lines in post-processing \cite{wesche_freedrawer_2001}, featuring the use of real physical objects in AR as guides \cite{reipschlager_designar_2019}, using haptic feedback of the controller as assistance \cite{elsayed_vrsketchpen_2020} or letting the user sketch on 2D surfaces and then transform the lines in 3D space \cite{arora_symbiosissketch_2018, jackson_lift-off_2016, wesche_freedrawer_2001, dorta_hyve-3d_2016}. Similar problems arise with digital sculpting, in which the user shapes desired forms out of primitives and pre-included models using virtual sculpting tools \cite{noauthor_transform_2020}. Other approaches include mesh modeling \cite{reipschlager_designar_2019, butterworth_3dm_1992} and assembly from primitives \cite{deering_holosketch_1995, reipschlager_designar_2019, do_3darmodeler_2010}.
Additionally, there also exist several commercially available tools for 3D design in VR that allow users to paint 3D artworks~\cite{noauthor_tilt_nodate}, design objects by free-hand drawing~\cite{noauthor_transform_2020}, and modify the mesh directly~\cite{toff_blocks_2017}.
Closely related to this paper, GetWild \cite{wong_getwild_2022} is a VR environment-editing system incorporating AI-generated models into the creation process to speed up the design pipeline and increase accessibility for users with little modeling experience. 

While allowing for expressive 3D modeling, direct modeling approaches can be complex for users with little modeling experience. 
In contrast, this paper investigates systems that co-creatively help users create 3D objects in VR.

\subsection{AI for 3D Object Generation and Manipulation}

While AI research for 2D image synthesis has made notable progress in the past few years, the field of 3D object generation has faced more difficulties due to limitations of the different representation modes of 3D data \cite{shi_deep_2023,li_generative_2023} as well as a lack of text-to-3D training data for deep learning models \cite{li_generative_2023}. Neural Radiance Fields (NeRF) \cite{mildenhall_nerf_2020} is a recently developed approach to representing 3D data, initially serving the purpose of 3D image reconstruction. Due to their ability to bridge the gap between 2D and 3D data, NeRFs can be combined with powerful text-to-image generative models to construct advanced text-to-3D generative models \cite{li_generative_2023}. This has led to a recent upsurge in AI tools for 3D object generation, such as DreamFusion \cite{poole_dreamfusion_2022} and Magic3D \cite{lin_magic3d_2023}. 
However, outputs of NeRF-based 3D generation are difficult to use in downstream 3D applications since a lot of 3D graphics software requires standard data representations like meshes \cite{li_generative_2023}. Shap-E \cite{jun_shap-e_2023} is a recent work demonstrating object generation based on a text prompt, with a mesh output created within less than two minutes.

Though generative models like the ones mentioned above are already helpful for exploring 3D design ideas, it can be difficult for users to formulate a textual prompt that will produce an output meeting all of the user's envisioned criteria. 
This one-step approach often forces the user to make the AI generate an entirely new object based on a rephrased prompt. 
For this reason, efforts have been made to develop AI tools that can modify a given 3D object according to a user's instructions.

SKED \cite{mikaeili_sked_2023} uses sketch-based instructions to enable interactive editing of 3D objects. 
In contrast, ShapeCrafter \cite{fu_shapecrafter_2023} generates a 3D object from a text-based description and lets the user refine object characteristics by recursively adding descriptive phrases.
Similarly, Text2Mesh \cite{michel_text2mesh_2021} can change the style of 3D objects: Given an input mesh and a text prompt, it modifies color and geometry details that adhere to the user-specified style. Additionally, SPAGHETTI \cite{hertz_spaghetti_2022} allows users to select parts of a 3D object and perform rough local transformations on them, based on which the generative model creates an edited version of the same object.

Although previous works include interactive elements similar to co-creative systems, none leverage the intuitive and expressive potential of spatial 3D interaction and presentation offered by VR. Instead, these systems rely on indirect instructions via text or 2D interface interactions. In contrast, this work focuses explicitly on VR environments.

\subsection{Co-creative Systems}

Human-computer co-creativity is a process in which a human and a computer contribute as equals to the same creative process \cite{davis_human-computer_2013}. 
Such co-creative systems have been commonly developed for various domains~\cite{rezwana_designing_2022}. Examples are game design~\cite{liapis_sentient_2013,lucas_stay_2017} and visual art~\cite{davis_empirically_2016,oh_i_2018,lin_it_2020}.

A stream of research has also started to investigate co-creative systems for editing 3D objects. For instance, Liu et al. \cite{liu_interactive_2018} have proposed a system where users can take turns creating and editing 3D objects with an AI.
Closely related, Calliope \cite{urban_davis_designing_2021} has explored interaction possibilities of using generative adversarial networks as an active collaborator in VR. However, it did not investigate the specific influences of different representation modes of such a collaborator on user perception.
Other than Calliope, co-creative systems are typically not studied in VR environments but in physical or digital 2D ones. We want to explore whether findings from those areas also apply to co-creative systems in VR.

\subsection{Embodied AI}
Embodiment can be defined in several ways, though Ziemke~\cite{ziemke_whats_2003} has conceptualized the most basic form of an embodied system as a system that can act on and be acted on by its environment. Other notions include physical embodiment, which necessitates a physical body, and organismoid embodiment, which further constrains the physical body to resemble a living being. Guckelsberger et al. \cite{guckelsberger_embodiment_2021} extend Ziemke's typology with the concept of virtual embodiment, which requires a virtual body that can act on and be acted on by a virtual environment.

Research has shown that the virtual embodiment of agents can improve motivation, positive attitude, and collaborative experiences in creative systems \cite{baylor_promoting_2009, rezwana_understanding_2022, davis_empirically_2016}. Additionally, organismoid embodiment enhances identification, empathy, and perceived creativity and is considered essential for co-creativity due to the unique perspective it provides \cite{guckelsberger_embodiment_2021}. Also, Kim et al. \cite{kim_does_2018} investigated the effect of embodiment on the perception of a digital AR assistant.
Moruzzi \cite{moruzzi_artificial_2022} found that AI embodiment increased perceived creativity in a collaborative artistic process, while Rezwana and Maher \cite{rezwana_user_2023} showed that AI personification influenced users' perceptions of the AI as an independent collaborator and affected their ethical views on ownership of co-created artifacts.

Studies on the effects of embodiment in VR have mainly examined AI agents without co-creativity. In contrast, studies on the embodiment of co-creative AI have mainly been carried out in non-VR environments.
We combine both promising directions of previous works by studying embodiment in the context of a co-creative object-generating agent embedded in VR.

\subsection{Non-player Characters in Games} %
In the domain of game development, the design of autonomous human-like agents, so-called Non-Player Characters (NPCs), was addressed in the context of believability \cite{10.1145/176789.176803}. Recent research explores the potential of NPCs in virtual reality~\cite{10.1145/3677098} and of LLM-based agents with more human-like reasoning, planning, and execution abilities~\cite{10.1145/3586183.3606763}. Moreover, in a systematic literature review,~\citet{wittmann2022} discuss NPC design patterns and how to transfer those to the design of AI systems for human-AI collaboration. The authors point out six relevant focus fields: NPC responsiveness, appearance, communication patterns, emotional aspects, behavioral characteristics, and player-NPC and NPC-NPC team structures.

Unlike NPCs, which are inherently embodied as avatars, co-creative tools for object building do not inherently require such embodiment. This property raises important questions about the role of embodiment in such tools and its interaction with factors like the highlighting and visualization of changes, which are the focus of this paper.

\input{figures/ModelingFlow5}

\subsection{AI Contribution Visualization}
Computational creativity can be assessed through the creator, the product, the process, or the environment, but humans often interpret a system's creativity based on appearance, behavior, and output \cite{rhodes_analysis_1961, lee_what_2020, parise_cooperating_1999, el-zanfaly_sand_2023, pearce_towards_2001}. Observing the creative process may lead to higher evaluations due to empathy and perceived effort, known as the "effort heuristic" \cite{colton_painting_2012, kruger_effort_2004}, and delays in AI output generation can increase user engagement and perceived control \cite{liu_how_2023}. However, empirical studies on incremental visualization of AI creativity are lacking \cite{linkola_how_2022}, prompting the need for research on both AI embodiment and perceptual evidence in co-creative VR systems.

Thus far, no empirical studies have investigated the effects of such an intentionally incremental visualization of AI contributions, especially not separately from the impact of AI embodiment. Therefore, we address this lack of research by contributing a study in which both embodiments and perceptual evidence in the form of incremental contribution visualization are examined in the context of a co-creative VR system.

%% file: figures/ModelingFlow5.tex
\begin{figure*}
 	    \begin{minipage}[t]{.19\linewidth}
		\centering
      	\includegraphics[width=\linewidth]{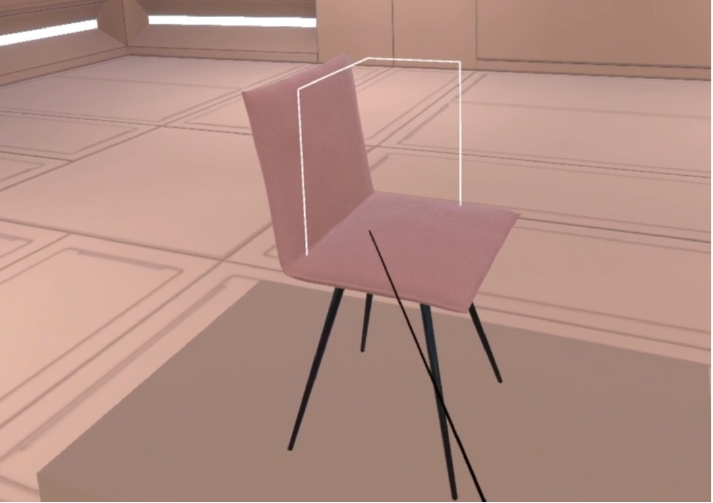}
		\subcaption{}
        \label{fig:results:ModelingFlow_a_Empty}
	\end{minipage}\hfill
  	\begin{minipage}[t]{.19\linewidth}
		\centering
      	\includegraphics[width=\linewidth]{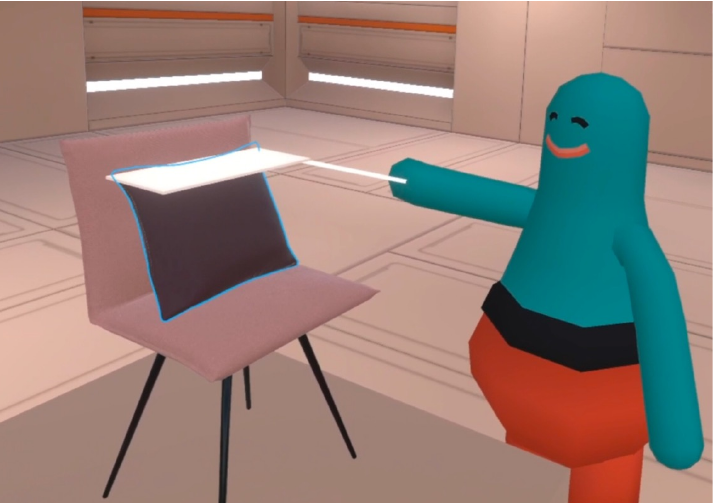}
		\subcaption{}
        \label{fig:results:ModelingFlow_b_AI_Creation}
	\end{minipage}\hfill
  	\begin{minipage}[t]{.19\linewidth}
		\centering
      	\includegraphics[width=\linewidth]{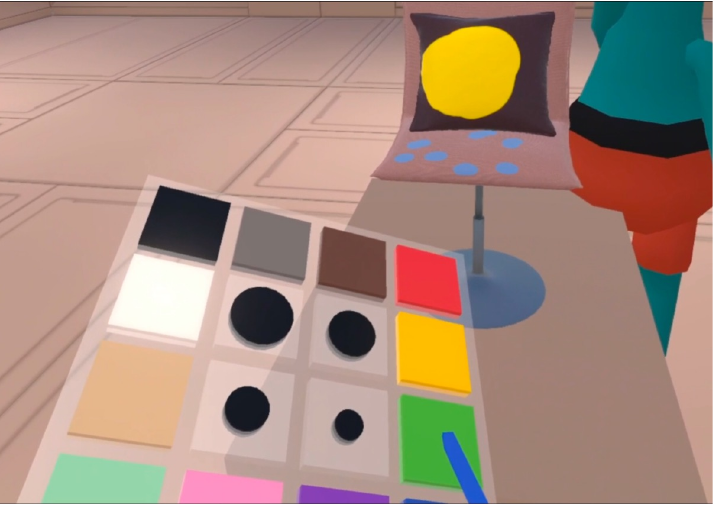}
		\subcaption{ }
        \label{fig:results:ModelingFlow_c_Painting}
	\end{minipage}\hfill
  	\begin{minipage}[t]{.19\linewidth}
		\centering
      	\includegraphics[width=\linewidth]{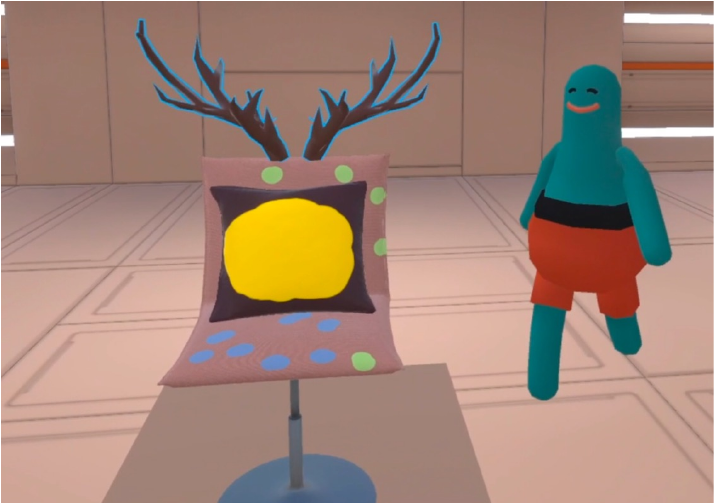}
		\subcaption{ }
        \label{fig:results:ModelingFlow_d_AI_Addition}
	\end{minipage}\hfill
  	\begin{minipage}[t]{.19\linewidth}
		\centering
      	\includegraphics[width=\linewidth]{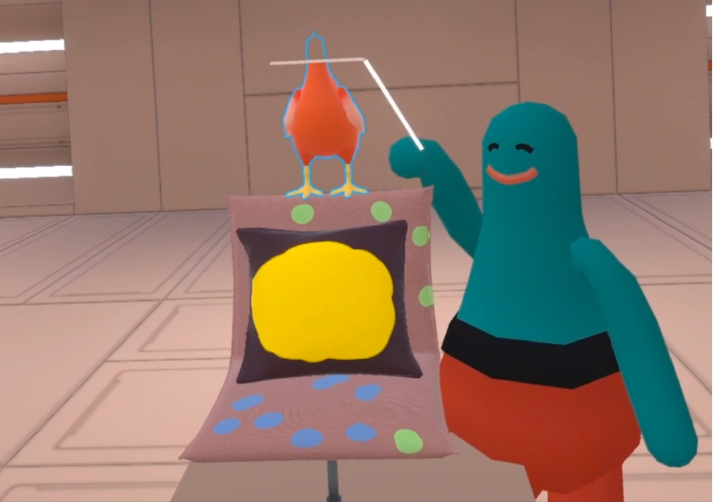}
		\subcaption{ }
        \label{fig:results:ModelingFlow_e_AI_Modification}
	\end{minipage}\hfill
	\caption{Steps during the collaboration creation at the example of a chair: (a) The chair at the center of the room on a podium (b) embodied, incremental, highlighted AI adds a new element to the selected position (c) the user paints elements using the palette, (d) AI adds another (creative) element to the chair (e) and modifies the previous addition.}%
 \Description{A chair in a VR scene with a hovering interface menu. An embodied AI that adds an object to the chair. A painting palette in VR with different colors and brush sizes. A modified chair with a yellow cushion and a pair of antlers. A chair with a chicken on top.}
	\label{fig:ModelingFlow}%
\end{figure*}

%% file: sections/method.tex
\section{Methodology}

Our study investigates the collaboration with an object-generating AI, focusing on the often overlooked crucial factor of AI-to-human communication~\cite{rezwana_designing_2022}.
We conducted a controlled experiment to examine how three representation modes (highlighting, incremental visualization, and embodiment) of the AI's generative contributions impact the co-creative process.
Our study is guided by the following research questions (RQs), each targeting one of the three representation modes:

\begin{description}
\item[RQ1] \emph{How does \textbf{highlighting} affect the co-creative collaboration experience, in particular regarding the predictability of AI?}   
Based on decades of HCI research and design practices that emphasize the importance of highlighting focus or activity, we hypothesize that highlighting will positively affect perceived communication quality, as it is a well-known mechanism to guide users' attention. 
For the same reason, we also expect that with highlighting, users will pay more attention to the AI, and the clarity of the AI intentions will be perceived better.
	
\item[RQ2] \emph{How does \textbf{incremental visualization} of changes affect the co-creative collaboration experience, in particular regarding a user's perception of the AI's efficiency and competence compared to a discrete visualization of the AI's contributions?}
Inspired by human collaboration, where it takes time for a modification to take shape, we expect that incremental visualization will affect measures concerning those same qualities as above, but also measures on the perceived system efficiency and competence, measures on the perception of AI outputs, perceived alignment of AI outputs with the users' visions, perceived agency, and the proportion to which contributions to the final artifact are attributed to the AI.
 
\item[RQ3] \emph{How does the \textbf{embodiment} of an AI affect the co-creative collaboration experience, in particular regarding the perceived supportiveness, efficiency, and competence of the AI?} 
Research on NPCs and games highlights embodiment as a critical factor in interaction, making embodiment a potentially promising factor for co-creative tools in VR. We thus expect that embodiment will affect the perceived supportiveness, efficiency, and competence of the AI, the perceived value and creativity of AI outputs, measures concerning collaborative experience, appeal, and behavioral engagement, as well as the previously mentioned contribution attribution and users' perceived closeness to the AI.
\end{description}

\begin{figure*}%
	\includegraphics[width=.9\linewidth]{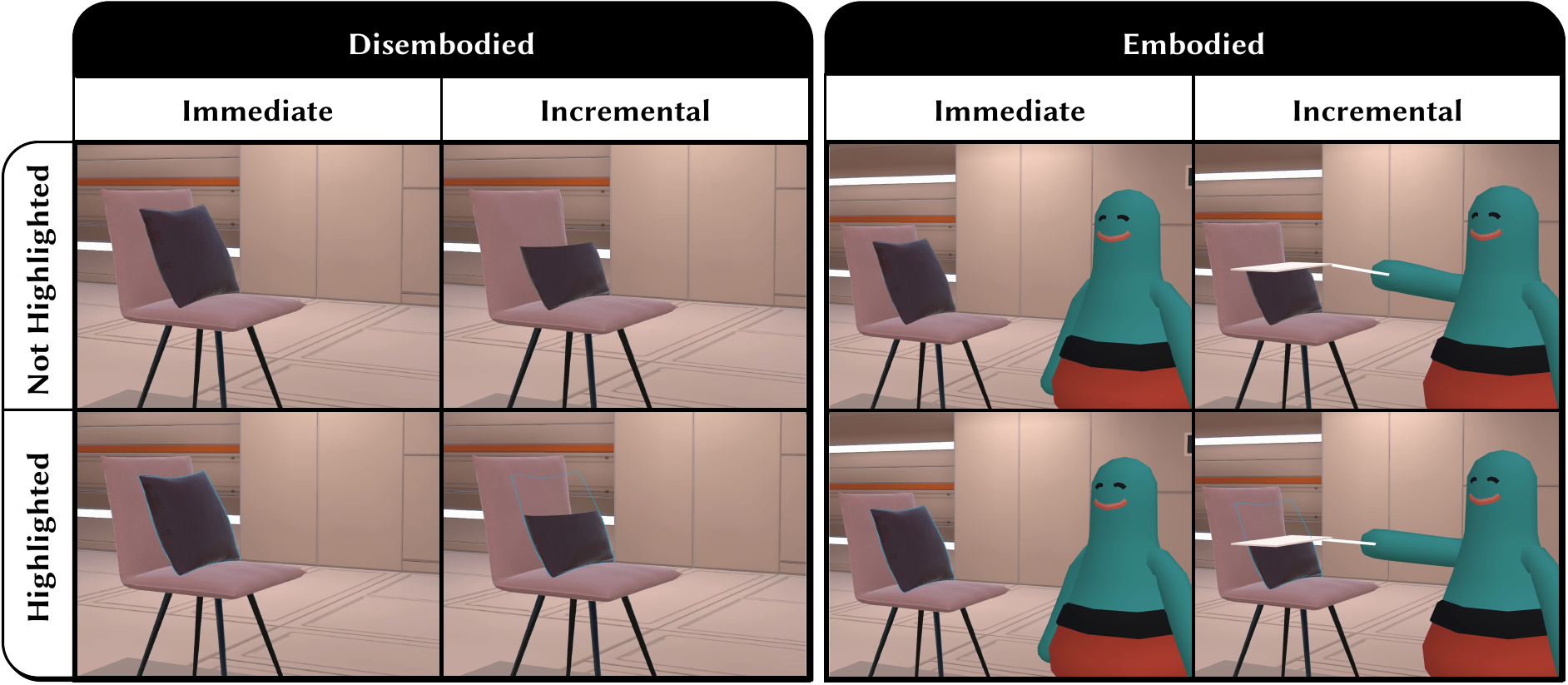}\vspace{.3cm} 
	\caption{We vary three different independent variables: (1) whether the AI has an embodied representation or is disembodied, (2) whether the AI's contributions are shown immediately or build up incrementally, and (3) if the area that the AI is going to modify is highlighted or not.}
 \Description{Showing the different IVs with an example of a chair. Left are four images illustrating the difference between disembodied AI. Right are four images showing the VR scene with an embodied AI}
	\label{fig:IVs}
\end{figure*}

\subsection{Task}
For our study, we placed participants in a virtual room, which was empty, besides a podium in the center (see \autoref{fig:results:ModelingFlow_a_Empty}) and a timer on the wall. In the embodied AI conditions, the AI avatar is shown additionally (see \autoref{fig:results:ModelingFlow_b_AI_Creation}). During the task, the user and the AI can collaboratively and iteratively modify a given 3D model. 

At the beginning of a modeling session, a 3D object appears on the podium, and the timer starts running. As long as the timer was running, the user could modify the object by either painting its texture (see \autoref{fig:results:ModelingFlow_c_Painting}), adding elements to the existing object (see \autoref{fig:results:ModelingFlow_d_AI_Addition}), or instructing the AI to modify (see \autoref{fig:results:ModelingFlow_e_AI_Modification}).
We chose painting as the users' primary design responsibility as it allowed for a consistent engagement with the 3D object during the AI-driven modifications and ensured comparable levels of object alteration and creative expression while avoiding interference.
For the modifications, users were always controlling if and when any changes to the object should be executed (including being able to revert the AI's modification).
We simulated the AI agent in a Wizard-of-Oz approach, in the sense that its contributions to the 3D objects were not the results of an actual generative process but instead object additions and replacements created by human 3D artists in advance. 

Further, the creative contributions of the AI agent are not designed to align with the contributions of the human partner but instead emerge completely independently of the human's behavior. This contribution style is typical for what Rezwana and Maher \cite{rezwana_designing_2022} refer to as \textit{provoking agents} (as opposed to \textit{pleasing agents}), which evoke creativity in their human partners by providing divergent ideas and alternative explorations.
While the AI was active, the user could not give the AI another modification instruction. %

We provided participants with four starting objects available for modification: A dinosaur, a chair, a pickup truck, and a house (see \autoref{fig:inflmodels}). %
Each object allowed for seven modifications randomly chosen by the system.
The AI modifications either consisted of an object part being replaced by another one (e.g., switching out the legs of a chair with legs of a different type) or an object part being added to the object (e.g., adding a dormer to the roof of the house).

While most modifications matched the object's type, a few unexpected changes were intentionally introduced (e.g., bunny ears on a dinosaur or cheese in a pickup truck). 
This design allowed us to explore how users perceive AI creativity, novelty, and predictability across different visualization settings.

\subsection{Design}

\subsubsection{Independent Variables}
In our study, we vary three representation modes that change how the AI carries out a modification instruction (see \autoref{fig:IVs}):

\begin{description}

\item[\ivHigh] 
If highlighting is enabled, instructing the AI to modify a part of the object will cause an outline to appear around the object part in question. If none of the other visualization aspects are enabled, this outline will be displayed for three seconds and then disappear at the same time that the object part is replaced with its modified version. If incremental visualization is enabled, the highlight will be displayed on the object part for exactly as long as the incremental process takes. If embodiment is enabled, the highlight will be displayed on the object part from the moment the AI avatar starts walking toward the object part until it has finished stepping back from the object. We base this concept of highlighting active elements on traditional 2D interface design to guide user attention.%

\item[\ivInc]
If incremental visualization is enabled, instructing the AI to modify an object part will cause this part to disappear and its modified version to slowly appear slice by slice as if a 3D printer was printing it. If the embodiment is enabled, a "printing layer" will be placed on top of the incrementally growing object part, vaguely resembling the additive manufacturing technique of sheet lamination. The AI avatar will appear to move this layer along the printing direction through a line connecting its right hand to one of the layer's corners.
	
\item[\ivEmb] 
If embodiment is enabled, an AI avatar will sit on a chair in the corner of the room at the beginning of a modeling session. Once instructed to modify a part of the 3D object, it will get up and walk toward the object part to perform the modification. If incremental visualization is enabled, it will stand beside the object holding the printing layer until the incremental process is finished. If incremental visualization is not disabled, the avatar will carry out a waving hand gesture reminiscent of a person performing a magic trick, after which the modified version of the object part replaces its previous version. Once the object part is finished being modified, the avatar will walk away a few steps from the object and wait in this location until it is instructed for another modification. After the AI has finished, it walks back to its chair autonomously.

\end{description}

 \begin{table*}[t!]
    \caption{The list of questions participants were asked in the semi-structured interview.}
	\resizebox{\textwidth}{!}{%
		\begin{tabular}{@{}ll@{}}
			\toprule
			Overall experience & How was your experience carrying out the given task? \\
			& Which of the runs did you prefer? \\
			& Which one did you like the least? \\ \midrule
			Perception of and satisfaction with the results & What do you think of your results? \\ 
			& Were you aiming for these kinds of results? \\ \midrule
			Perception of AI & How would you describe the relationship between you and the AI in the context of these tasks? \\
			& Would you have preferred the Ai to do more/less/other things? \\
			& In an optimal future version of this system, how would the AI behave? \\ \midrule
			Suggestions & Do you have any other comments or suggestions? \\ \bottomrule
		\end{tabular}%
	}
	\label{tab:interviewqs}
\end{table*}

When none of the representation modes were enabled, the modification process solely consisted of the object part being replaced by its modified version instantaneously without highlighting or any additional embodiment.

We varied all three independent variables in a repeated measures design, resulting in a total of $2 \times 2 \times 2 = 8$ conditions. %
We counterbalanced the order of all factors in a Balanced Latin Square to prevent learning effects.
Each participant conducted eight modeling trials across eight different conditions. In each session, participants modified one of the four 3D objects (see \autoref{fig:modmodels}). Each object was presented to each participant twice, with the order randomized for each individual. 
We chose four objects as a suitable compromise, striking a balance between variety and familiarity. 
This approach allowed participants to engage their creativity by attempting various design approaches with the same model while also providing a second chance to revisit and revise previous modifications – an opportunity to learn from failure that we argue is valuable in our research context.

\subsubsection{Dependent Variables}
Data collected during the study consisted of the following:

\newcommand{\paragraphB}[1]{{\textbf{#1}}}
\paragraphB{Before the experiment:} Demographic information was collected through a Google Forms survey.

\paragraphB{During modeling sessions:} A log of user interactions with the system was recorded. This consisted of the type of each interaction with the object and the AI, as well as the time at which it was performed. Additionally, a screencast of the participant's perspective in VR was recorded in each modeling session.

\paragraphB{Between modeling sessions:} After each modeling session, the user was asked to fill out a questionnaire.
We base our questionnaire on several other questionnaires from the HCI and Psychology:
\begin{itemize}
	\item The Mixed-Initiative Creativity Support Index (MICSI) \cite{lawton_drawing_2023} measures the degree to which a co-creative system supports a user in creative tasks. For the purposes of our questionnaire, we adopted eight items, some of them as slightly modified versions: the two items on \textit{Enjoyment} and one item each on \textit{Alignment} and \textit{Agency} – all measured on 7-point Likert scales – and one item each on \textit{Contribution}, \textit{Satisfaction}, \textit{Surprise}, and \textit{novelty} – all measured on 7-point non-Likert scales.
	\item Rezwana and Maher \cite{rezwana_understanding_2022} modified the Creativity Support Index (CSI) \cite{cherry_quantifying_2014}, which the MICSI was based on, to include two items evaluating \textit{partnership} and \emph{communication} in human-AI collaboration. We adopted both of those items in our questionnaire.
	\item The User Experience Questionnaire (UEQ) \cite{laugwitz_construction_2008} measures general user experience for interactive systems and consists of items on seven-point adjective scales. We adopted five of those items: One from the \textit{efficiency} scale, two from the \textit{Dependability} scale, and two from the \textit{Stimulation} scale.
	\item The Inclusion of Other in the Self (IOS) Scale \cite{aron_inclusion_1992} is a pictorial single-item scale measuring \textit{closeness} between the respondent and another person. Though the AI in our co-creative system is not a person, we are nevertheless interested in whether closeness, when applied to a co-creative AI partner, is affected by our visualization aspects. We, therefore, included this item in our questionnaire.
	\item The Social Presence in Gaming Questionnaire (SPGQ) \cite{de_kort_digital_2007} was developed for digital games and measures players' awareness of and involvement with their co-players. Though our co-creative system is not a game, we are interested in whether our visualization aspects influence the degree to which users perceive a co-creative AI as a separate presence that influences and can be influenced by the user. This is why we adopted all eight items from the \textit{Behavioral Engagement Scale} of this questionnaire, all of which are answered with 5-point intensity scales.
\end{itemize}
In addition to the items sourced from the above-mentioned questionnaires, we included three Likert items on AI competence, contribution value, and AI creativity, amounting to 27 questions in total\footnote{See the supplementary materials for a list of all items}.
 
\paragraphB{At the end of the experiment:} In a semi-structured interview, participants answered questions about their overall experience with the system and their perception of the finished 3D objects and the AI (see \autoref{tab:interviewqs}).

\subsection{Apparatus}
The study was carried out using a Windows 11 computer with an NVIDIA GeForce GTX 1080 graphics card and a Meta Quest Pro head-mounted display (HMD) connected to the computer via Quest Link. The software was run directly from its Unity environment. Participants were given both of the Meta Quest Pro controllers and were free to walk around in a room-scale tracked space. Questionnaire replies were filled out on a separate computer.

\input{figures/quant_data_plot}

\subsection{Procedure}
After welcoming participants and before starting the study, participants filled out an informed consent form and completed a demographic information survey. We then provided an overview of the study procedure and system usage guidelines.
Next, we instructed the participants on using the HMD and controllers and let them familiarize themselves with the VR interactions in a virtual room by testing selection and painting operations on a sphere. 
Once ready, participants began their first modeling session, tasked to \textit{"be creative and work together to create a new version of this model"}, emphasizing a focus on making the given object look different from its default state but letting participants choose their own design goals.
We timed a 5-minute session (visualized by a virtual clock) and automatically disabled all interactions once time ran out. 
Participants were invited to voice any thoughts they had during the sessions.
Participants then filled out questionnaires outside the VR environment before repeating this process for a total of eight conditions. After completing all sessions, we conducted a semi-structured interview with each participant. 
In total, the study lasted around 90 minutes. All participants were given cake as compensation for their time.

\subsection{Participants}
We recruited 16 participants (7 identified as female, 9 as male, and 0 as non-binary). Ages ranged from 20 to 30, with a mean age of 24 ($SD=3$). 
Out of all the participants, five had no prior experience with VR, while nine had tried it before, and two had used it more often. 
Nine participants had no experience with 3D modeling, three had tried it before, and four had done it more often. 
Most participants had interacted with AI in the form of Large Language Models before. In contrast, half of the participants had used AI for image generation before, and only three participants had experience with AI for 3D object generation. 

%% file: figures/quant_data_plot.tex
\begin{figure*}[t!]
	\includegraphics[width=\linewidth]{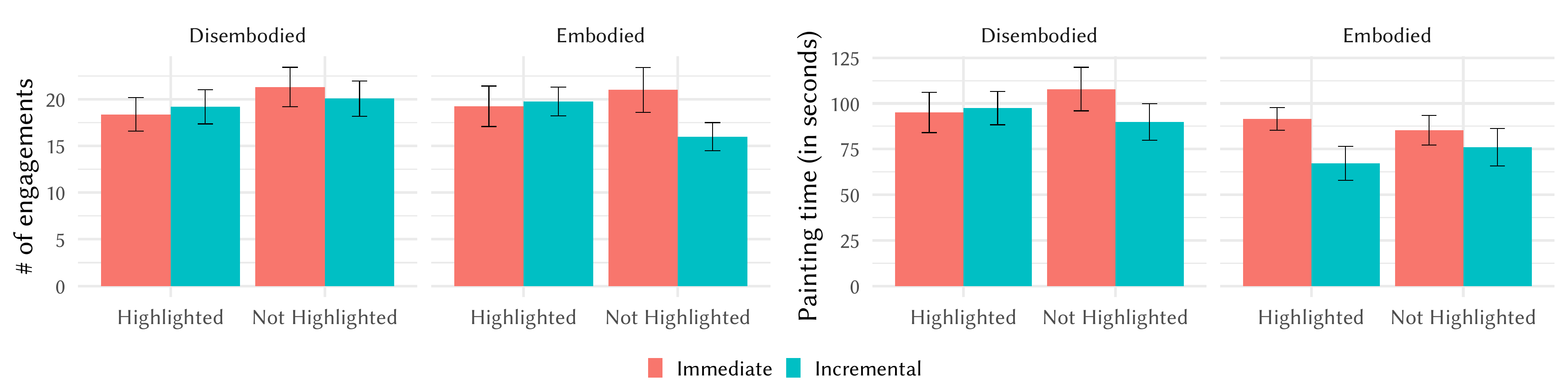}
	
	\begin{minipage}[t]{.49\linewidth}
		\centering
		\subcaption{Number of Engagements}\label{fig:results:nr_of_engagements}
	\end{minipage}%
	\begin{minipage}[t]{.49\linewidth}
		\centering
		\subcaption{Painting Time}\label{fig:results:painting_time}
	\end{minipage}%
	
	\caption{The number of engagements and painting time in our experiment. All error bars depict the standard error.}
 \Description{Bar plots of number of engagements and painting time with error bars.}
	\label{fig:results:walking}
\end{figure*}

%% file: sections/results.tex
\section{Results}

\begin{figure*}
	\centering
    \includegraphics[width=\linewidth]{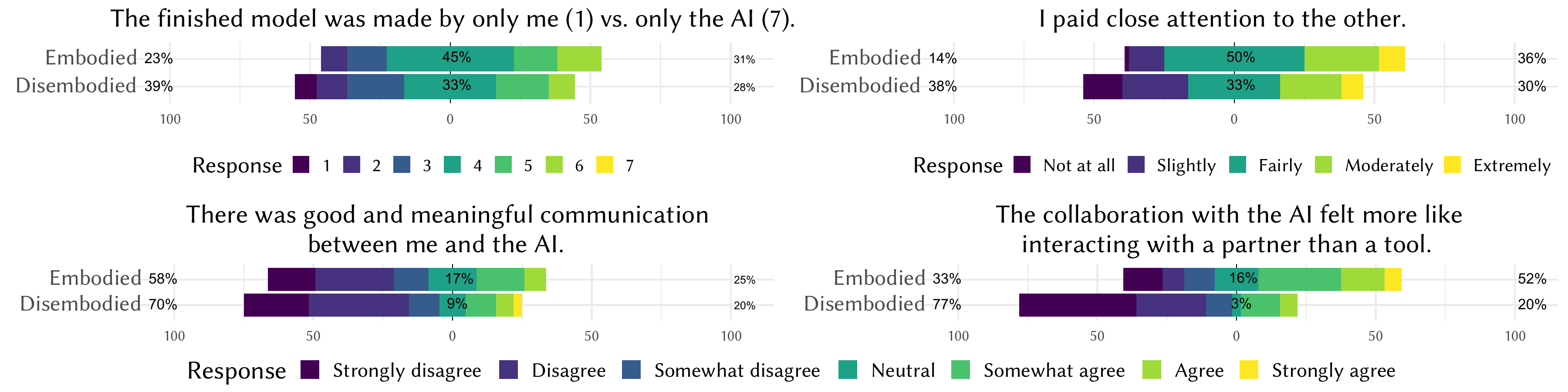}	
	\caption{Significant effects of embodiment on attributing contributions to the AI, perceived communication and partnership, and attention paid toward the AI. The plots show the percentages of participants who gave a response to a question, grouped by whether embodiment was enabled in the corresponding modeling session or not.}
 \Description{Likert scale plots for the questionnaires}
	\label{fig:embodiment-plots}
\end{figure*}

In this section, we report our empirical study's quantitative and qualitative results.

\subsection{Quantitative Results}

We analyzed the influence of our independent variables on the dependent variables using inferential statistics. 

\subsubsection{Number of Engagements}

We analyzed the number of engagements to measure the participant's interest in contributing to the model (see \autoref{fig:results:nr_of_engagements}). We defined the number of engagements as the total number of selections, modifications, and undo actions that a user performed. We found results ranging from \val{16.0}{6.0} (with \ivHigh{} and \ivInc{} without \ivEmb{}) to \val{21.3}{8.4} (without any ai contribution visualization).

For statistical analysis, we fitted Poisson
regression models and applied Type III Wald chi-square tests for significance testing. The analysis indicated a significant (\chisq{1}{6.29}{<.05}) main effect of \ivHigh{} on the number of engagements that was, however, not supported by post-hoc tests.

Further, we found a significant (\chisq{1}{6.68}{<.01}) interaction effect between \ivHigh{} and \ivInc{}. Without \ivHigh{}, the addition of an \ivInc{} significantly ($<.05$) reduced the number of engagements. With \ivHigh{}, however, the addition of \ivInc{} seemingly increased the number of engagements ($n.s.$)

Finally, we found a significant (\chisq{1}{4.86}{<.05}) interaction effect between \ivHigh{} and \ivEmb{}. Here, we found the same trend where without \ivHigh{}, \ivEmb{} reduced the number of engagements while increasing it with \ivHigh{}. However, these results were not significant.

\subsubsection{Painting Time}

As another measure for the engagement of the participants, we measured the time participants spent painting parts of the model (see \autoref{fig:results:nr_of_engagements}). We found painting times ranging from \valS{67.2 }{37.1} (with all ai contribution visualizations) to \valS{108.0}{47.8} (without any ai contribution visualization).

We analyzed the data using 3-way repeated measures ANOVA. We tested the data for violations of normality and sphericity assumptions using Shapiro-Wilk's and Mauchly's tests and found no violations.

The analysis indicated a significant (\anova{1}{15}{4.926}{<.05}) main effect of the \ivInc{} on the painting time. Post-hoc tests confirmed significantly ($p<.05$) higher painting times without the \ivInc{}.

Further, the analysis indicated a significant (\anova{1}{15}{9.349}{<.01}) main effect of the \ivEmb{} on the painting time. Again, post-hoc tests confirmed significantly ($p<.01$) higher painting times without the \ivEmb{}.

\subsection{Questionnaire Results}
We analyzed the responses of our participants using the Aligned-Rank Transform (ART) procedure as proposed by \citet{wobbrock_aligned_2011} followed by ART-C~\cite{elkin2021} corrected using Tukey's method for post-hoc comparisons where appropriate. 
\autoref{fig:embodiment-plots} depicts some of the most interesting effects found in our analysis.

\subsubsection{Perception of AI}
We assess the perception of AI in general by analyzing the questions on predictability, support, efficiency, and competence.
None of the visualization aspects significantly affected predictability, efficiency, or competence. However, in conditions with embodiment, participants rated the system as significantly more supportive ($F(1, 15)=5.74$, $p<.05$).

\subsubsection{Perception of AI output}
We assess the perception of AI output by analyzing the questions on unexpectedness, unusualness, value, and creativity.
There were no significant differences in ratings for unusualness, value, or creativity for any visualization aspects. However, in conditions with incremental visualization, participants rated the finished 3D model as significantly more unexpected ($F(1, 15)=7.52$, $p<.05$).

\subsubsection{Collaborative experience}
We assess the collaborative experience by analyzing the partnership, communication, alignment, and agency questions.
There were no significant differences in ratings for alignment or agency. However, with embodiment, participants rated communication significantly higher ($F(1, 15)=26.19$, $p<.001$). Additionally, participants rated partnership lower when highlighting was enabled ($F(1, 15)=6.18$, $p<.05$) and higher when embodiment was enabled ($F(1, 15)=23.5$, $p<.001$). We also found a significant interaction of highlighting and embodiment concerning this measure ($F(1, 15)=11.02$, $p<.01$). Post-hoc tests do not show significant differences in ratings between the individual variable combinations.

\subsubsection{Appeal}
We assess the overall system appeal by analyzing the questions on excitement, motivation, affinity, and enjoyment.
There were no significant differences in ratings for motivation and affinity. However, in conditions with highlighting, participants enjoyed using the system significantly less ($F(1, 15)=7.89$, $p<.05$). Additionally, we found a significant interaction of highlighting and embodiment concerning the measure of excitement ($F(1, 15)=5.83$, $p<.05$). Post-hoc tests do not show significant differences in ratings between the individual variable combinations.

\subsubsection{Behavioral engagement}
We assess behavioral engagement by analyzing the questions from the Behavioral Engagement Scale.
Only two questions showed significant differences between the enablement of any visualization aspects: The extent to which the AI affected what the participant did was larger when incremental visualization was enabled ($F(1, 15)=6.54$, $p<.05$), and the amount of attention participants paid to the AI was more significant with embodiment ($F(1, 15)=9.43$, $p<.01$). We also found a three-way interaction of all visualization aspects concerning the latter variable ($F(1, 15)=7.87$, $p<.05$); however, post-hoc tests do not show significant differences in ratings between the individual variable combinations.

\subsubsection{Closeness, satisfaction, and contribution}
We assess these factors by analyzing satisfaction, contribution, and closeness questions.
There were no significant differences in closeness to the AI as measured by the IOS Scale. However, in conditions with highlighting, participants were significantly less satisfied with the finished 3D model ($F(1, 15)=5.06$, $p<.05$). In conditions with embodiment, participants attributed significantly more of the contribution to the finished 3D model to the AI ($F(1, 15)=9.19$, $p<.01$).

\subsection{Qualitative Results}
\subsubsection{Data  Analysis}
We collected qualitative data in a final semi-structured interview.
Our data analysis procedure is based on Braun and Clarke's \cite{braun_using_2006} step-by-step approach of thematic analysis.
While two authors conducted the analysis, the results and intermediate steps were discussed among all authors.
The analysis was performed based on the transcripts of the audio-recorded interviews. The interview responses were a mix of English and German. All quotes were translated into English by the authors, who are fluent in English and German.
The two authors first familiarized themselves with the data by reading the transcripts. 
We used Delvetool~\footnote{Delvetool for qualitative data analysis \url{https://app.delvetool.com/}, last accessed: August 30, 2024} for coding the data.
In the first analysis step, both authors read the same subset of transcripts (8 transcripts/50\%) separately and created initial codes inductively. They then discussed the codes and formed themes by grouping them using an online mind map tool~\footnote{Miro tool for online mind maps: \url{https://miro.com/}, last accessed: August 30, 2024}. After the initial theme generation, all authors discussed the themes before one of the authors coded the remaining transcripts using the updated codes. Some new codes were merged with the existing themes during the second coding stage. Finally, the two authors met again to discuss the final codes and update and name the themes.

\begin{figure*}
	\centering
		\subfloat[\centering Dinosaur]{{\includegraphics[width=3.6cm]{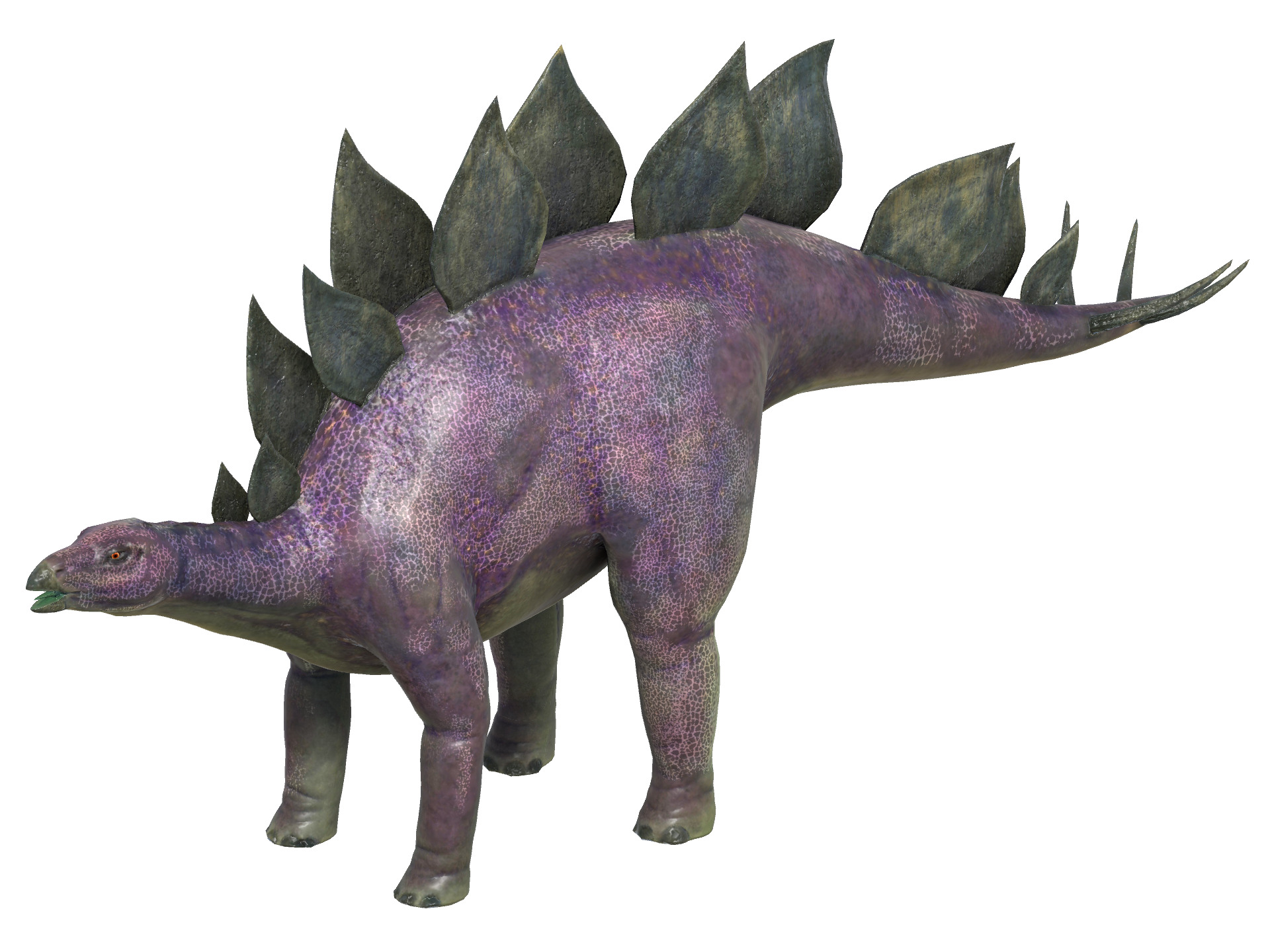} }}\vspace{.07cm}
		\subfloat[\centering Pickup]{{\includegraphics[width=3.6cm]{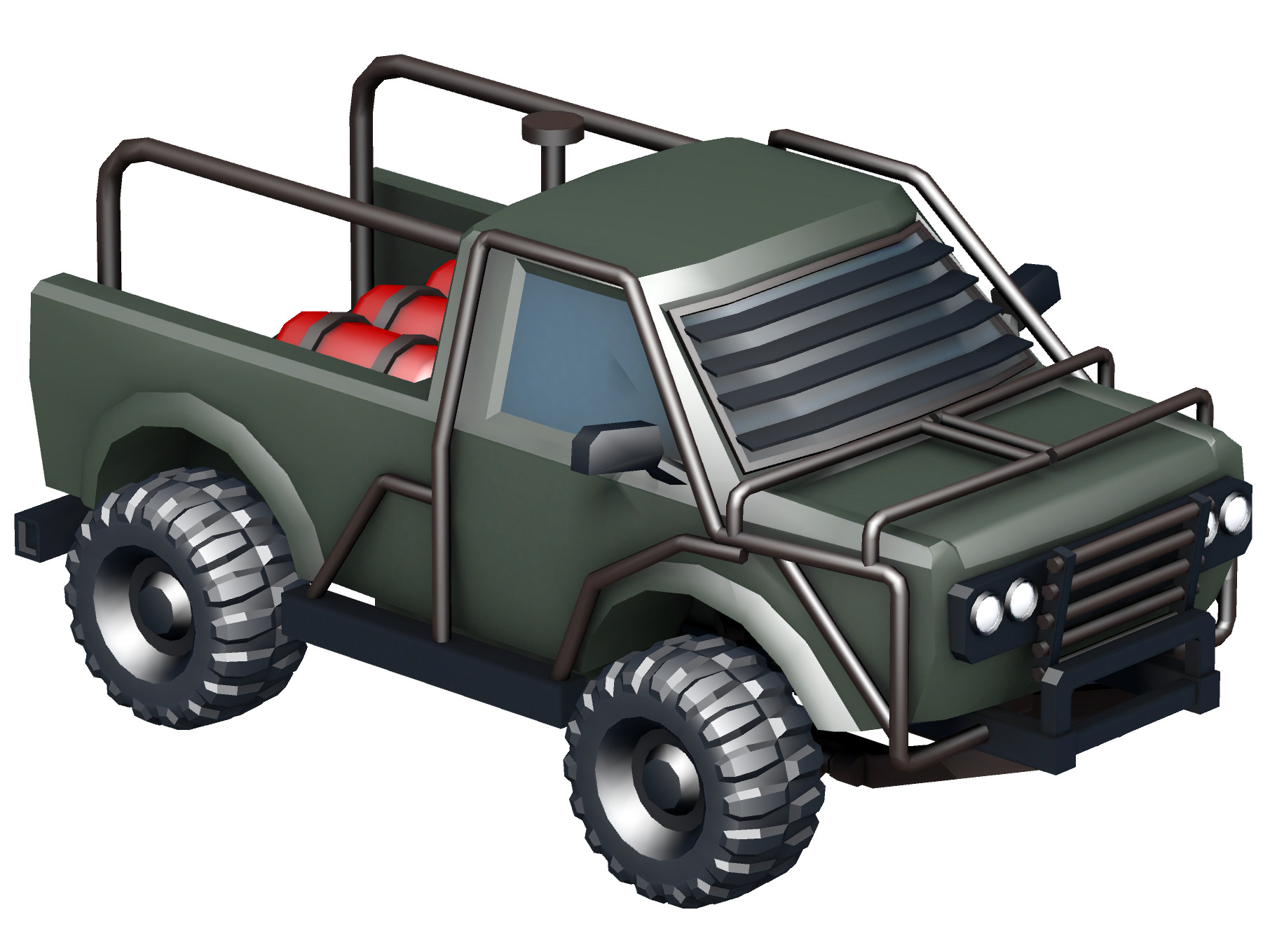} }}\vspace{.07cm}
		\subfloat[\centering Chair]{{\includegraphics[width=3.6cm]{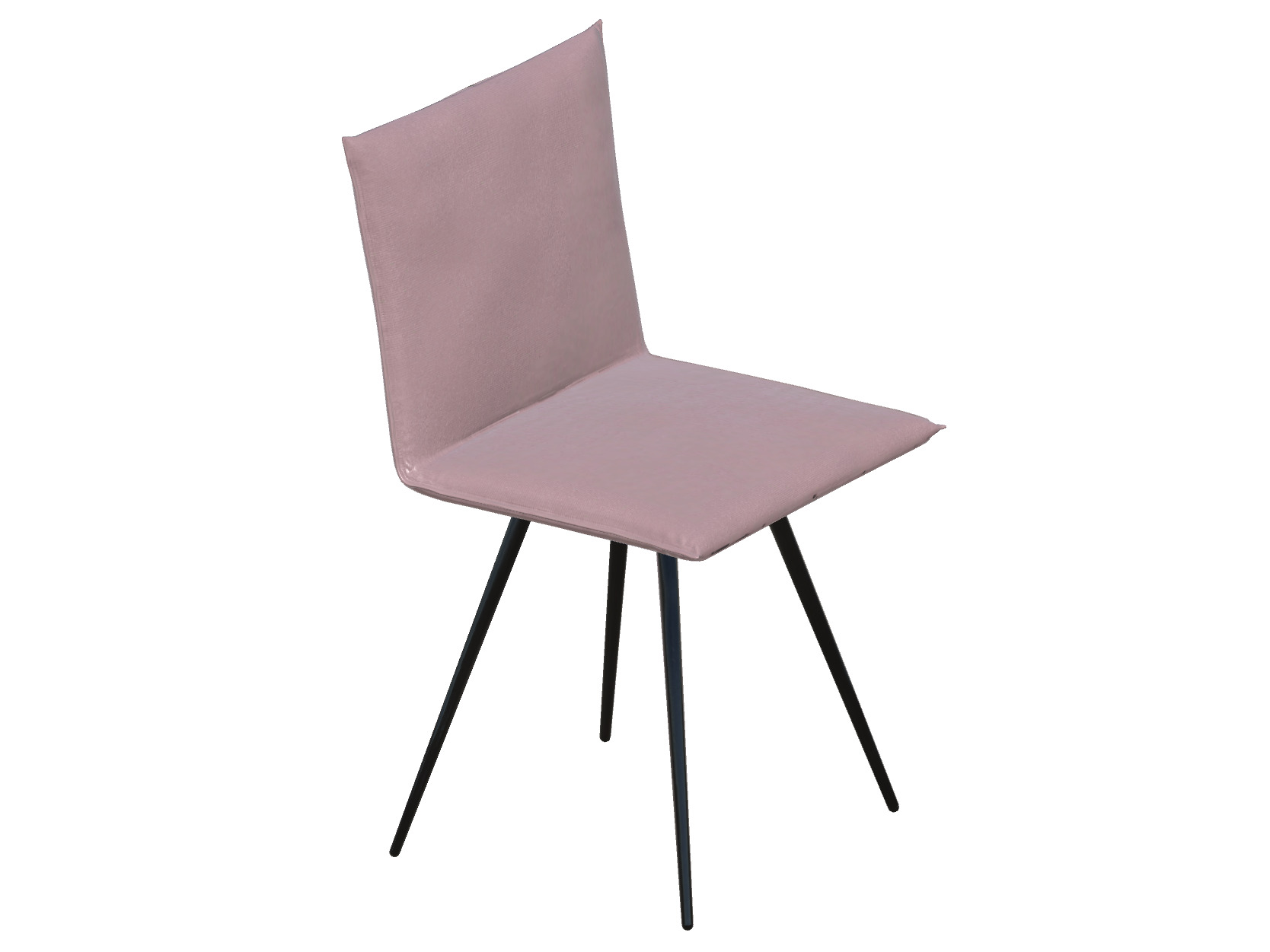} }}\vspace{.07cm}
		\subfloat[\centering House]{{\includegraphics[width=3.6cm]{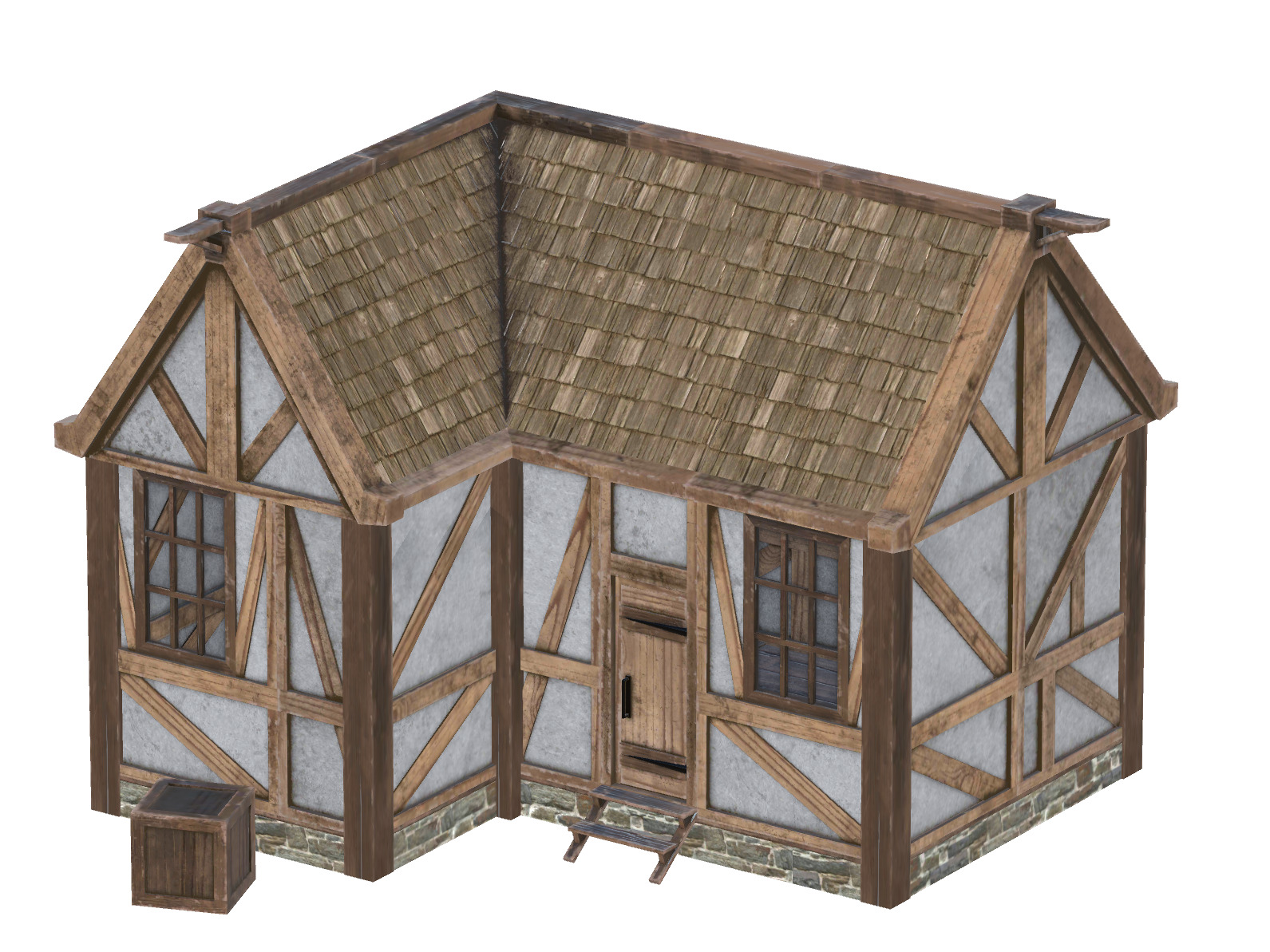} }}\vspace{.3cm}
	\caption{The four modifiable objects in their default form before any modifications.}%
 \Description{Four 3D objects: A dinosaur, a pickup, a chair, a house}
	\label{fig:modmodels}%
\end{figure*}

\begin{figure*}
	\centering
	\includegraphics[width=.85\linewidth]{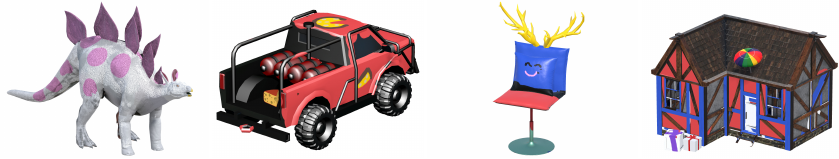}		
	\caption{A selection of finished models designed by participants, each influenced differently by the AI or its contributions.}
 \Description{Four modified objects: A purple dinosaur, a colorful pickup, a colored chair with a pair of antlers, and a colorful house.}
	\label{fig:inflmodels}
\end{figure*}

\subsubsection{Codes and Themes}
Coding the first half of the transcripts produced 103 codes.
They were then grouped into 21 initial summary codes.
Before coding the second half of the transcripts, we discussed and summarised the codes into 73 unique codes.
Next, the second half of the transcript was coded, during which we added 15 codes to the coding set and merged them with the existing ones into 20 summary codes. We then formed the following four themes based on the summary codes\footnote{See the mind map of the codes in the supplementary materials}.

\subsubsection{Task Division and Control}
\textbf{Participants experienced varying levels of control over the AI, ranging from dependency to independence and different hierarchical levels.}
One key aspect discussed by participants was how to divide tasks between themselves and the AI, particularly concerning the level of control each would have (or desire to have) over the task.
Their comments revealed a relationship axis involving a degree of dependence/independence and, in the case of dependence, a hierarchy of interaction. The participants' comments were quite diverse, showing no standard agreement on the type of control they felt over the AI.
Some participants (P13, P15) mentioned they did not feel they were working with the AI. Others noted that they felt like they were co-working with the AI but not collaborating (P8): \textit{"It felt more like ... when you're [..] doing exercise sheets, and you have a partner who never talks to you and just hands something in without you even noticing." }. Some participants also stated that they ignored the input from the AI (P15).
Although the AI's input was indeed predetermined, some users felt in control of its actions and described a hierarchical relation (P1). They viewed the AI as a tool rather than a collaborator or co-worker. In this context, participants felt that the AI should support the user.
The third relationship we observed was that of a partner, where participants felt they were pursuing a common goal with the AI.
While many participants did not note any particular influences of embodiment on the relationship, P5 explained that they felt that collaboration was more relevant to them than working with an embodied AI.
Finally, some participants described feeling that the AI was in control and that their decisions were overruled by it (P2, P8).

Participants also wished for different types of relationships when asked for improvements or desired changes in the interaction.
Some participants suggested a form of task sharing with the AI, where both would choose from a set of options (P4).
Other participants desired a more hierarchical way of control, where they wanted to give commands to the AI (P1, P2), while others were okay with the AI working on its own as long as the user could make the final decision about accepting or rejecting its decisions (P2): \textit{So, [the AI] should be able to try out things, but in any case, it should not override my decisions.}
These wishes are closely related to the desire for better communication between AI and the user (P3, P6). For example, the AI should indicate whether they understood the user's request and whether the requested action was possible (P3). P2 also indicated they would like to receive feedback from the AI about its perception of the task.

In addition to spoken or verbal feedback, participants indicated that they would like to interact with the AI through different modalities and meta-level instructions, for example, by pointing at things or by making general style requests that are then performed by the AI (P2: \textit{"for example I can write: I want to make a goth house, and then the AI automatically suggests goth windows and so on.}).

\subsubsection{AI as Enabler of New Experiences}
\textbf{Interacting with the AI sparked creativity among participants and enabled new experiences.}
We observed that some participants felt the AI enabled them to have more creative experiences. For example, participants liked the AI's suggested modifications (P1). P8 liked that these suggestions guided them towards a satisfying result, and P7 appreciated that the AI suggestions led to new inputs: \textit{[...] so when they [the AI] conjured something new, then, of course, one got new input.} P2 also mentioned that they felt inspired by the AI: \textit{[...] the things from the AI also inspired me during the process.} Some participants even mentioned that they felt the AI helped them to be more creative and open (P7, P8), stating that they were happy with the result, although they, for example, chose a color scheme that they would not have gone for themselves. %

\subsubsection{Emotional Experience with AI} %
\textbf{Participants had both positive and negative experiences with the AI.}
Participants generally found the experience interesting (P2, P6, P7).
Their opinions and experiences with the AI were both positive and negative. On the positive side, participants felt that the AI created funny results and a funny experience in general (P3, P4). On the negative side, participants mentioned that they were confused about the AI and its created outcome (P2). They also did not like the slow response time or waiting for the AI to finish their modifications (P4, P5). Finally, some participants felt surprised (P3, P4) by the suggestions or distracted by the AI output (P1, P3).

\subsubsection{Experiences with Embodied AI}
\textbf{The embodied AI evoked a variety of emotions.}
The participants experienced quite diverse emotions when interacting with the embodied AI. Some found it annoying because they had to wait for the embodied AI to finish the animation (P5) and disliked it being slow (P3, P4, P5, P7). While some participants liked to watch the AI, others felt it was distracting (P3). However, participants also felt curiosity when working with the embodied AI (P8). 

Despite these negative feelings, participants also expressed positive emotions towards the AI. For example, they perceived it to be helpful (P1), cute (P2), or funny (P2). Participants also made comments about the appearance of the embodied AI. Some participants found it uncanny and weird (P3). Some participants specifically criticized the avatar's permanent smiling expression, explaining that \textit{"smiling for no reason is creepy"} (P11) and mentioned that they did not prefer the AI to be human-like (P7). Indeed, several participants noted that they would like to be able to customize the appearance of the embodied AI (P3). P5 found it even scary when seeing the AI for the first time: \textit{"Yes, I would say that during the iterations, I got used to it, but all in all, I was quite scared when I saw it the first time"}. However, we also observed positive interaction effects. P1 stated that \textit{"changes felt more tangible when done by the embodied AI"}. P4 felt that social conventions applied to the interaction with the AI: \textit{"it was nicer to wait when at least there was this person who made something"}, and P7 stated that the embodied AI helped not to feel alone. Overall, some participants felt a connection to the embodied AI. P7 even said they missed the embodied AI when it disappeared for some of the iterations. Finally, while some participants did not like the slow version of the continuous interaction mode, they felt it was more acceptable when the AI was embodied versus when it was not.

\subsection{Participants' Approach and Task Limitations}
This section summarizes how participants generally approached the task. It also describes the limitations introduced by the study task unrelated to the human-AI interaction.

Almost all participants divided their modeling sessions into two stages: First, they searched for AI modifications and let the AI perform all the options they could find, undoing any modifications they were not happy with. The second stage was typically spent entirely on painting. In a few cases, participants were satisfied with the model before the five minutes had elapsed and spent the remaining time mostly just looking at their creations. Most participants did not discover all possible modifications that could be performed on the four models.

Participants often criticized the fact that only limited modifications were possible for the given models and that potential changes did not vary when a model appeared for a second time. Some participants hoped to be able to influence the AI's modifications through their behavior. P11 voiced the name of the object that he was hoping to appear in a modification spot. At the same time, some participants tried to elicit specific modifications using the colors they painted with, and P8 even wrote a modification prompt on the object's surface itself.

Participants were also bothered that it could not be re-instantiated once they had undone a modification. They moreover criticized several aspects of the painting mechanism, such as the limited amount of colors, the lack of an eraser, and the issue that if a model had repeating textures, painting on one part of a model could make a copy of the painted stroke appear at another location on the model that used the same texture.

Some exemplary images of finished models created by participants during the modeling sessions are depicted in \autoref{fig:inflmodels}.

%% file: sections/discussion.tex
\section{Discussion and Implications}
In this section, we revisit our quantitative and qualitative analysis results to interpret them and derive implications for the future design of co-creative VR systems. Additionally, we discuss the generalizability and limitations of this work.

\subsection{Highlighting does not alleviate the confusion caused by AI actions}
We found the highlighting to decrease perceived partnership, enjoyment, and satisfaction with the finished model. At the same time, highlighting did not affect the system's predictability, the perceived communication between humans and AI, the users' attention toward the AI, or the clarity of AI intent, which we expected to be most likely affected by highlighting. On the qualitative side, the observational analysis also did not indicate that highlighting object parts caused differences in participant behavior. This result is supported by participants never referring to highlighting during the modeling sessions or interviews. 
Our quantitative and qualitative findings suggest that highlighting is not as much of a mitigator against user confusion about AI intentions as expected and might lead to a more negative overall experience. Future research should investigate whether these adverse effects can be replicated or if they occur similarly with different implementations of a highlighting variable.

\begin{leftbar}{lightgray}
    \noindent \textbf{Design implication for highlighting}\quad
    For co-creation use cases where users create one complete object at a time, like in our use case, we recommend not highlighting changes. In particular, in combination with embodied AI avatars, it introduced more negative effects than it contributed to user satisfaction. Future work has to show if highlighting proves more useful for co-creation use cases of dynamic objects.
\end{leftbar}

\subsection{Incremental visualization increases the attention to AI actions but introduces unexpected side effects}
Since most participants watched all incremental visualization processes from beginning to end, it is unsurprising that this variable significantly affected the attention participants paid to the AI. However, we also expected that the visualization aspect would affect the perceived efficiency of the system, which our quantitative analysis did not confirm. Considering that participants were likely aware that the AI was not generating model parts "on the spot," they likely interpreted this process as the loading time of the overall setup. This interpretation also explains why the incremental visualization did not affect predictability, competence, unusualness, value, creativity, or contribution, as we hypothesized.

Further, this lack of impact on the perception of the AI and its outputs could be caused by the specific incremental visualization we used. This incremental visualization was inspired by AI drawing agents that build images stroke-by-stroke. Transferring this concept directly into 3D space resulted in an AI agent that slowly draws individual horizontal lines of different colored pixels, which does not correspond to our experiences in the physical world. A possibly more appropriate 3D equivalent could be the molding or carving of primitives.
Future work should investigate whether alternative ways of incrementally visualizing 3D modifications affect observers' perceptions of the system differently.

Another surprising finding is that incremental visualization increased the perceived unexpectedness of AI outputs. This finding contrasts with our hypothesis that incremental visualization would lead to a decrease in unexpectedness rather than an increase, as the longer visualization duration gave participants more time to familiarize themselves with the modified version of an object part. We attribute this reverse effect to participants having had more time to engage with the version of an object part due to the extra time (e.g., to think about what the fully instantiated version of the object part would look like or what kind of change they would have expected instead at this point). As a result, they perceived the AI outputs as more unexpected than in conditions where the output appeared immediately, and participants had no choice but to immediately accept them as the new version of an object part.

Additionally, our interviews revealed that incremental visualization was perceived more positively when an avatar appeared to carry out the incremental process. We attribute this to a more sympathetic attitude towards an AI with characteristics similar to the user's. An AI that does not have a physical or virtual body might be expected not to be limited by time and space to the same degree as a human and might, therefore, seem more capable of performing actions that would be impossible for a human user. On the other hand, an embodied AI might appear to be constrained by its surroundings in similar ways to a human. It would make sense for a human user to be more sympathetic toward AI, which "has to" slowly work on its task until it is finished, and therefore, to feel more like the AI is equal to the user. One participant also applied this idea to the "tool" that the AI used in the incremental process, which was a line that looked similar to the pointer ray that participants used to interact with the VR environment. It would be interesting for future work to study whether the effects of this similarity of tools only apply when a similarity of processes (i.e., an incremental process) is also given or whether the same effect can be observed given an instantaneous visualization.

\begin{leftbar}{lightgray}
    \noindent \textbf{Design implication for incremental visualization}\quad
    We recommend using an immediate creation of content to co-create complete objects, where the creation process is unimportant. We ground this in our findings where incremental generation did not reduce participants' feelings of unexpectedness, while at the same time, the visualization took longer. However, we recommend incremental visualization for use cases where the process of creating objects is important and where users should observe the generation more closely, and time is no relevant factor.
\end{leftbar}  

\subsection{Embodiment not only affects the relationship to the AI but also to the 3D model created}
Embodiment was the visualization aspect that participants had the most substantial feelings about. The divergence in comfort levels regarding the presence of the avatar was unexpected: To avoid an avatar design that created feelings of uncanniness in users~\cite{mori_uncanny_2012}, we intentionally chose an avatar for our AI that was heavily caricatured and only distantly resembled the form of a human, as well as one that had a friendly, child-like appearance. We do not know whether participants' discomfort with the avatar was indeed a case of the uncanny valley, whether it was only due to the surprise participants felt when they were faced with the avatar for the first time, or whether it can be attributed to characteristics that are specific to co-creative systems. 

The divergence in participants' discomfort with the avatar is also a possible explanation for why embodiment did not significantly affect ratings of factors measuring appeal. Nevertheless, our qualitative analysis indicated an improvement in the perceived relationship between users and the AI when the avatar was present, manifested by an increased tangibility of the relationship, the feeling of being less alone, and more empathy expressed towards the AI. This result is consistent with quantitative data pointing to an effect of embodiment on partnership and AI-human communication.

As hypothesized, AI embodiment did affect the extent to which participants felt that the AI contributed to the finished 3D model. This result suggests that factors external to the AI functionality can manipulate users' perceptions of who created an artifact. This finding is noteworthy considering ongoing discussions around ownership of AI-generated content, as the question of who created an artifact is often one of the most influential in deciding who owns it \cite{rezwana_user_2023}. Rezwana and Maher \cite{rezwana_user_2023} found that this decision of ownership, and additionally users' stances on other AI-related moral dilemmas, is also greatly influenced by the perception of a co-creative AI partner as either a collaborator or as a tool. Considering our finding that the embodiment of our AI significantly affected participants' perceptions of the AI as a partner rather than a tool, we argue that such ethical questions should be kept in mind when deciding whether to integrate an AI avatar into a co-creative system. This result is in line with Buschek et al. \cite{buschek_nine_2021} listing unclarity of ownership as a potential pitfall in the design of co-creative systems.

\begin{leftbar}{lightgray}
    \noindent \textbf{Design implication for embodiment}\quad
    We recommend using an embodied AI for most co-creation use cases, as it increases the perceived support of the system, leads to a higher perceived partnership and communication, and leads to users paying more attention to the creation. However, users also felt reduced ownership of the outcome, which should be considered when designing a system. For use cases where users need a tool and the creative, collaborative aspect is less relevant, the non-embodied AI might be the appropriate choice, as it is faster, and users have higher ownership of the outcome.
\end{leftbar}

%% file: sections/limitations.tex
\section{Limitations and Future Work}
Our work provides valuable insights and implications for the future design of co-creative systems for object generation in VR. 
In particular, we are confident that our technology-agnostic approach, which does not rely on current AI capabilities or technologies, will ensure that the results remain relevant for future systems, even if these are not constrained by the same technological limitations. %

However, we acknowledge that our experiment focused on \textbf{specific AI representation modes}, and other variants or visualizations may yield different results. 
We selected these specific modes of representation because of their importance for human collaboration to explore their impact on users' perceptions of the co-creative process and to stimulate further research and alternative implementations.

Moreover, the \textbf{limited set of models} that participants were asked to modify was an extraneous variable, potentially affecting participants' experience with and perception of the system. Though we tried to minimize the effects of this variable by randomizing the order in which models were presented to each participant, we expect that with a small sample size such as ours, effects could not be avoided.

Also, the \textbf{limited number of possible AI modifications}, which remained constant during iterative modifications, could have led participants to the conclusion that their AI partner did not actually generate object parts but retrieved them from previously modeled parts. We hypothesize that this may have had an impact on the perception of AI and its results, as there was nothing about AI's capabilities and how its "generative" process changed between conditions.

Our study also did not explore the potential of a co-creative AI in \textbf{more complex, object-building scenarios}, such as simultaneous adjustments to multiple copies of the same object, adding and manipulating entire environments, or deformation and distortions of the object mesh (cf. \cite{Slim2024,Achlioptas2023,Angelis2024} on 3D shape editing and deforming).
We deliberately opted for a constrained set of predefined actions to focus our contribution on the fundamental questions of co-creative interaction and to provide a better understanding of co-creative tools in VR for object-building that are more than one-shot tools. While we are convinced that this provides a robust foundation for future work, we acknowledge that the constrained set of predefined actions might limit the generalizability of our findings, and future work in this domain is necessary.
Exploring how the AI's functionality scales with increased diversity and complexity of tasks is an important direction for future research, particularly for understanding the full potential of AI in co-creative VR systems.

Moreover, the efficiency of the co-creative tool itself was not explicitly evaluated in our study. 
The results presented thus may be correlated with how effectively the co-creative system executed its tasks, potentially influencing users' perceptions of the system's creativity. 
As generative AI becomes more advanced and prevalent in co-creative processes, future research should examine the role of AI efficiency and its impact on users' experiences and outcomes in co-creative tasks. Such investigations could provide a deeper understanding of how system performance and responsiveness affect the perceived value and usability of co-creative systems.

Since we performed the study as a controlled experiment, we acknowledge limitations in \textbf{external validity} in favor of establishing cause-effect relationships to recognize the impacts of individual representation modes. %

There are several interesting directions for \textbf{future work}:
One important area is exploring how different methods of emphasizing the AI's contribution process might positively impact a user's experience with a co-creative system, including whether such methods can enhance user navigation and coordination during collaboration. 
Additionally, investigating various ways to incrementally visualize the AI's generation of a 3D object could reveal how these visualizations affect users' perceptions of the AI and its creative process. 
There is also a need to examine the specific characteristics of avatars for co-creative AI that may cause discomfort to users, as well as how the perceived closeness to AI agents is affected by their embodiment and which other factors might influence this perception. 
Finally, future research should also explore whether manipulating the visual appearance of an AI's embodiment can alter people's understanding and evaluation of the AI and its creations.

%% file: sections/conclusion.tex
\section{Conclusion}
Research on co-creative systems has predominantly focused on how humans participate in the co-creative process, often overlooking how AI contributions are represented and perceived by users. 
This paper addresses this gap by examining how an AI's contributions can be effectively represented for co-creative object generation in VR.
Through a Wizard-of-Oz study in a VR-based co-creative 3D object-generating environment, we investigated the effects on the co-creative collaboration experience for three key representation modes AI can use to communicate to the user: highlighting, incremental visualization, and embodiment of AI contributions.
Finally, we derive implications for the design of tools for co-creative object-building: First, designers should avoid combining highlighting with embodiment to increase user satisfaction. Second, while an embodiment increases perceived support, it also reduces ownership of the created objects. Third, designers should favor immediate visualization when the creation process is secondary.

%% file: meta/acknowledgments.tex
\begin{acks}
We would like to thank all participants and reviewers. 
This research was supported by the HumanE AI Network from the European Union’s Horizon 2020 research and innovation program under grant agreement No 952026, and the Pioneer Centre for AI, DNRF grant number P1.
\end{acks}

%% file: bib/bibliography.tex
\bibliographystyle{bib/ACM-Reference-Format}
\bibliography{bib/references}